\documentclass[12pt]{iopart}
\pdfoutput=1
\usepackage[utf8]{inputenc}
\usepackage[english]{babel}
\usepackage[T1]{fontenc}
\usepackage{hyperref}

\usepackage{tikz}
\usepackage{lipsum}
\usepackage{subcaption}

\usepackage{amssymb}
\usepackage{epsfig,pstricks,graphicx}
\usepackage{bm}
\usepackage[shortlabels]{enumitem}
\usepackage{ulem}
\expandafter\let\csname equation*\endcsname\relax
\expandafter\let\csname endequation*\endcsname\relax
\usepackage{color}
\usepackage{array}
\def\id{{\rm 1\kern-.22em l}}
\usepackage{mathtools}
\usepackage{braket}

\renewcommand\ket[1]{{|{#1}\rangle}}

\usepackage[format=plain,justification=centerlast]{caption}
\usepackage[format=plain,justification=centerlast]{subcaption}
\usepackage{dsfont}

\usepackage{iopams}  
\begin{document}

\title[]{On physicality of electromagnetic potential from causal structure of  flux quantization}
\author{Konrad Schlichtholz\footnote{Corresponding author}}
\address{International Centre for Theory of Quantum Technologies (ICTQT),
University of Gdansk, 80-309 Gdansk, Poland}
\ead{konrad.schlichtholz@ug.edu.pl}
\author{Marcin Markiewicz}
\address{International Centre for Theory of Quantum Technologies (ICTQT),
University of Gdansk, 80-309 Gdansk, Poland}
\address{Institute of Theoretical and Applied Informatics, Polish Academy of Sciences, ul. Baltycka 5, 44-100 Gliwice, Poland}
\ead{marcin.markiewicz@ug.edu.pl}

\vspace{10pt}

\begin{abstract}
Recent work by Vaidman [Phys. Rev. A 86,
040101 (2012)] showed that Aharonov-Bohm effect can be explained in terms of  local fields, thus effectively restating an old problem of physicality of potentials. In this work, we propose an argument demonstrating the physicality of electromagnetic potential (upon the assumption of locality) based on the causal structure in flux quantization setup. Crucially, we discuss the fundamental difference between the considered setup and the Aharonov-Bohm experiment that allows for avoiding Vaidman's loophole in our scenario.
\end{abstract}
      
%\maketitle

\maketitle
\section{Introduction}\label{sec1}

\subsection{The nature of electromagnetic potential -- theoretical perspective}
The dispute about whether electromagnetic potentials should be considered physical after decades still has not reached its end. This problem originates in the strong opinion of a part of the scientific community that objects treated as physical entities should be assigned values that are in principle uniquely determined through measurements. In classical electrodynamics, this opinion is only a preference and does not pose any problems.  This is because Maxwell equations can be fully stated using unique vector fields $\vec E$ and $\vec H$ or non-unique potentials $\vec A$ and $\phi$ that yield all the same predictions. Potentials corresponding to given fields are defined up to gauge transformations and thus rather define an abstraction class yielding a given prediction of theory and are often treated simply as a mathematical tool only. Although one may prefer to avoid more involved mathematical structures in the fundamental formulation of the theory keeping it as simple as possible, one should not forget that it is only a mathematical construction that tries to mimic our observations. If all predictions of two structures are the same, one is unable to falsify the hypothesis that one is more fundamental than another (one is physical and another is not). Here, more fundamental should be interpreted as more resembling physical reality, which we try to model, as in the end it could turn out to be different from both those structures. 

However, this discussion could not be hopeless as we go from classical to quantum theory. In quantum electrodynamics, charged particles are coupled with an electric field through minimal coupling. This coupling involves potentials instead of fields in order to retrieve the classical equations of motion for charged particles. The typical argument for that is that if Hamiltonian (or equivalently Lagrangian), which is the starting point for quantum theory, were to contain fields directly, the Hamiltonian or Lagrangian densities corresponding to interactions would contain the derivatives of the fields. This would imply that the interactions would be of shorter range than the inverse-square-type ones, which is not the case for the electromagnetic field. (see \cite{Weinberg1}, Chapter 5.9).

Nevertheless, the status of electromagnetic potentials in the logical structure of constructing quantum field theory (QFT) remains unclear for another reason. In this construction, on the one hand, we have single-particle states corresponding to irreducible representations of the Poincare group (Wigner's classification); on the other hand, we have fields with well-defined Lorentz-transformation properties, the quanta of which should correspond to the particles. Now, photons in the Wigner classification have helicities equal to $\pm 1$, which gives rise to their transversal two-dimensional polarization. There should exist corresponding spin-$1$ field of which they are excitations. Natural choice for such field seems to be the quantized electromagnetic four-potential $A_{\mu}=(\tfrac{1}{c}\phi, \vec A)$, however such straightforward identification is impossible for two reasons \cite{Weinberg1, BBBB}: (i) the quantized four-potential should have $A_0=0$ in all Lorentzian frames due to non-existence of time-like photons, which is clearly impossible for any four vector field, (ii) the commutation properties of $A_{\mu}$ are inconsistent with the commutation properties of the fields, if fields are directly canonically quantized, and the standard relation between potentials and fields is assumed. What is important both the above difficulties hold independently of the choice of the gauge (even if the gauge is Lorentz covariant) \cite{BBBB}. Therefore, the quantized electromagnetic potential within quantum field theory must be treated in a non-covariant frame-dependent  way, which somehow spoils the mere logic of the construction. 

On the other hand, as mentioned earlier, it seems necessary to be involved in describing interactions between fields and sources. This necessity is supported by the gauge-theoretical description of fundamental interactions, in which electromagnetism arises as a factor restoring invariance of the Dirac field representing electrons and positrons with respect to local $U(1)$ transformations. In this approach electromagnetic potential plays the role of a \textit{connection} field, which enables defining covariant derivative for the Dirac field. This connection is specified up to gauge transformations, nevertheless its presence and coupling with the four-current is a necessary element of restoring gauge invariance of the theory (at the same time justifying necessity of the minimal coupling description of interaction between fields and charges). At the same time the electromagnetic field tensor also naturally appears in this framework and plays the role of a curvature tensor.

To sum up, the status of the electromagnetic potential within the quantum field theory is also ambiguous, and one has to search for more operational arguments to resolve the issue of its "physicality".

\subsection{The nature of electromagnetic potential -- operational perspective}

Since electromagnetic potential can be non trivial in the region where the field is zero, one might find that a charged particle could be in some way influenced in the absence of fields. With the assumption of locality of interactions, this would indicate that particles interacted with the potential, thus casting it as physical and more fundamental than fields. Such a reasoning has led to the discovery of the Aharonov-Bohm (AB) effect first described in \cite{Ehrenberg49} and then rediscovered by Aharonov and Bohm \cite{AB59} in the context of showing the physicality of the potentials. This famous effect describes the phase difference gained by the two beams of charged particles going around an infinite solenoid with some magnetic field flux $\Phi$ present  inside. As the field is confined to the solenoid, the phase difference is associated with the interaction with the potential $\vec{A}$.

Although the existence of the effect was long debated, see e.g. the following works suggesting its non-existence via introducing non-standard forms of a vector potential \cite{Bocchieri78, Bocchieri80, Roy80} \footnote{Such potentials, which on the one hand assure constant field inside the solenoid, on the other hand do not lead to the AB effect, contain singularities and  violate Stokes theorem. As suggested by some authors, such potentials should not be allowed in description of physical reality, see e.g. \cite{Klein79, Greenberger81}.}, it finally found multiple experimental confirmations \cite{Chambers60, Tonomura86, Osakabe86, Webb85}. 
It is worth mentioning, that there exists a different approach to understanding the AB effect which emphasizes that the effect has purely topological origin, and therefore it goes beyond the ''fields vs. potentials`` dispute. Namely it states that the effect can be explained by noting that the vacuum of the experiment, understood operationally as the region of the configuration space in which the energy density of the fields is zero, is not simply connected, as is the case of $\mathbb R^3$ with removed infinite cylinder $\mathbb R\times S^1$, see  \cite{Wu75}, \cite{Ryder} (sec. 3.4). 
Although such geometrical configuration is indeed characteristic to AB-type arguments, the notion of \textit{vacuum} in this argument does not adequately correspond to the understanding of vacuum in QFT (problem of \textit{zero-point energy}), therefore it cannot be treated as a complete physical justification of the effect.

Let us note that in the literature the time-dependent version of the AB effect is also considered, where the flux in the solenoid varies in time and thus generates non-zero field outside the solenoid. However, the effect of time-dependent flux is still highly debated \cite{TD1,TD2,TD3,TD4,TD5,wakamatsu2024revisitingcontroversytimedependentaharonovbohm} and does not have proper experimental confirmation.

The experimental verification of the standard static AB effect did not end the debate of whether the AB effect certifies the physicality of potentials or the presence of some nonlocal interactions. 
Finally, the work by Vaidman \cite{Vaidman12}  provided what we call Vaidman's loophole, namely an explanation of the AB effect in its basic setup in terms of locally interacting fields (this was also more rigorously analyzed by removing some semiclassical approximations in \cite{Pearle1,Pearle2}). This explanation is based on the generation of an entanglement of a superposed charged particle with the solenoid by the magnetic field generated by the charged particle moving around the solenoid. Although Aharonov \cite{Aharonov15} disputed whether this explanation could be used for all modified versions of the setup, some of the proposed counterexamples were contested \cite{Vaidman15}. Therefore, while this result does not show a loophole that is certainly present in all modifications of the AB effect, it sheds doubt on whether one is able to argue the physicality of the potentials based on some setup concerning the AB effect. Still, the topic of AB effect finds broad attention, for example, with recent interest in the context of the locality of acquiring the phase \cite{Vedral20, Saldanha2021,Kang22,AB_shielding} on nonclosed loops. However, this problem is not directly related to the physicality of potentials. This is because, these papers using different methods show that second quantization predicts the phase difference at any point of a path around a solenoid which is acquired due to the local interaction with photons arising from quantization of the electromagnetic potentials, but they do not in any way try to show that those mediating photons do not contribute to a  non-zero field, e.g. field generated by the moving electron which is crucial in Vaidman's loophole.

Since the AB effect may never give us the final conclusion after all in the debate of \textit{potentials versus fields}, one might shift the attention to different effects. Note that here we refer to the AB effect specifically as the effect of emergence of phase difference in the recalled experiment above and not in a more general context of geometrical phase \cite{berry1984quantal,Cohen2019} in  electrodynamics. 

Another well-known effect, which has topological connotations analogous to the AB effect, is the flux quantization in superconducting rings \cite{london1950superfluids}. One of the principal equations in the theory of superconductors is the phenomenological London equation \cite{London1935TheEE}, which states that  in the superconductor there appears an induced current that is proportional to the vector potential $\vec A$. This equation in London gauge can be written as $\vec j=-\frac{n_s(e^*)^2}{m^*}\vec A$, where $\vec j$ is a current density in the superconductor, $e^*$ stands for charge of charge carrier, $n_s$ is the charge carriers density in superconductor and $m^*$ is the mass of the charge carrier. This equation was also shown to be consistent with the predictions of the microscopic BCS theory of superconductivity \cite{BCS}. Aside from the Meissner effect, the primary consequence of this equation is flux quantization. Having a superconductor with the hole one can find that the flux of a magnetic field through this hole is quantized with flux quanta $\Phi_Q=\pi\hbar /e$ where $e$ stands for elementary charge. This phenomenon has a wide experimental confirmation \cite{Flux_q_exp1,Flux_q_exp2} including quantization of the magnetic field in a long solenoid going through the superconducting ring \cite{DEAVER1982178} in the context of dependence on winding number.  In this context it is clear that a theoretical setup consisting of a superconducting ring with infinite solenoid inside could bear analogous conclusions to the AB effect due to the effect of flux quantization. This is because the flux quantization effect relies on the appearance of the current, which by the London equation can appear also in the region of zero electromagnetic field. However, such considerations have not attracted much attention in the context of the physicality of potentials \cite{Rev_flux} compared to the AB effect itself.

The declared non-physicality of potentials is even discussed along with the London equation in textbooks \cite{combescot2022superconductivity}, as Meissner effect could be derived solely from the version of the London equation that includes only fields ($\nabla \times \vec j=-\frac{n_s(e^*)^2}{m^*}\vec B$). Recently, the article \cite{Kenawy18} analyzed the time evolution of induced supercurrents by vector potential and the resulting flux quantization in such a setup. The results even more directly suggest AB-effect-like conclusions on the physicality of potentials by disjointing the Meissner effect from the flux quantization. Still, this analysis leaves some space for loopholes in the context of showing the physicality of potential as, for example, it does not address the field appearing outside the solenoid when the current is introduced into the solenoid and does not address in any way Vaidman's loophole.

In this paper, we discuss a gedanken experiment on modification of the setup concerning flux quantization, which suggests physicality of electromagnetic potentials,by which we mean a feature that an occurrence of a  measureable physical effect (here appearance of a supercurrent in a specified time window) is necessarily mediated by the sole presence of the electromagnetic potential in some region of space. Crucially, we also discuss the key difference between the AB effect and the setup considered, which allows one to avoid  Vaidman's loophole. This suggests that shifting research attention from the AB effect towards flux quantization in the context of the physicality of potentials might be a good direction for future research.

\section{Results and Discussion}
\label{sec2}
\begin{figure}
    \centering
    \includegraphics[width=\linewidth]{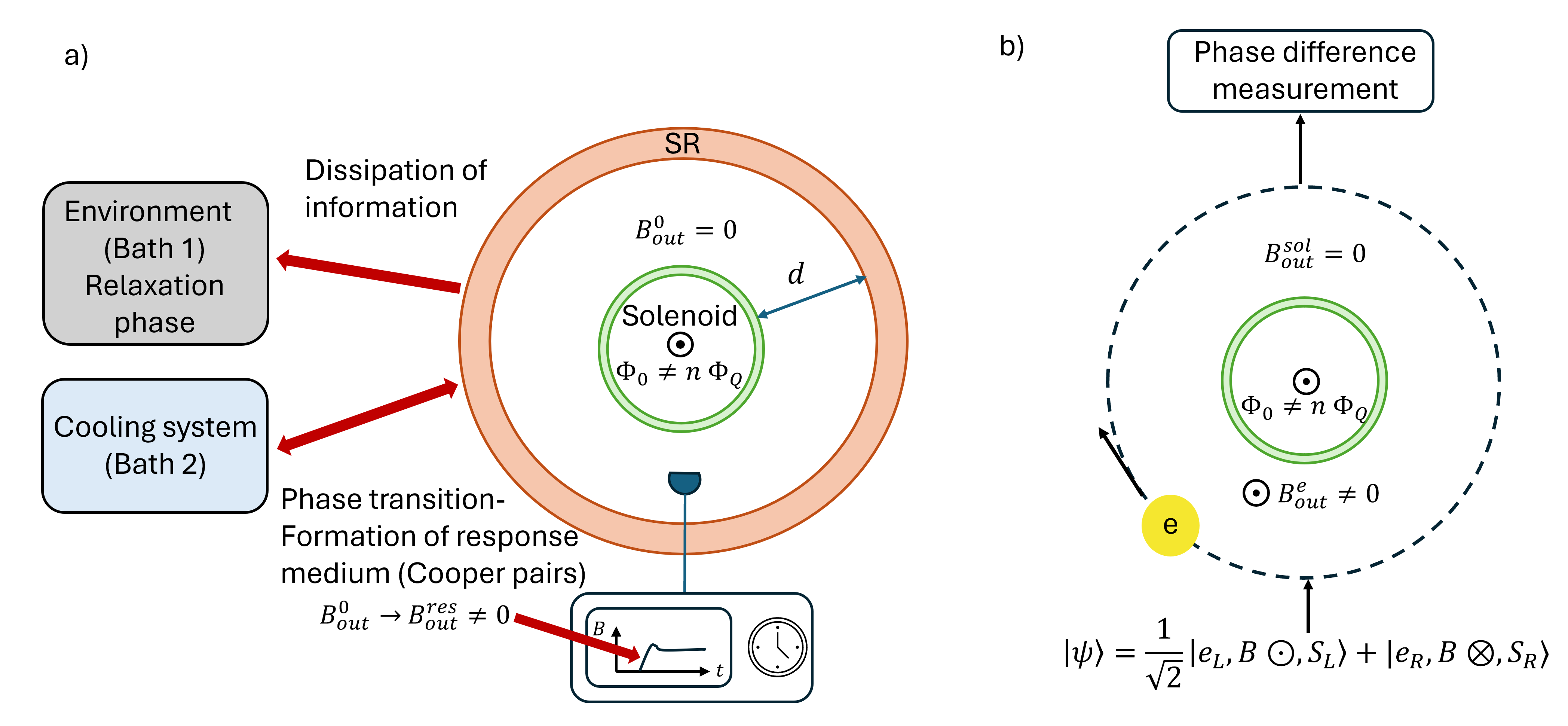}
    \caption{a) Scheme of the setup for the gedanken experiment. After introducing the flux not equal to the integer multiple of the flux quantum to the solenoid, the superconducting ring (SR), initially prepared in the normal (non-superconducting) phase, dissipates the information about the field appearing during the initial flux generation into the environment. Then SR is coupled to a cooling system to induce phase-transition which results in appearance of supercurrent in response to magnetic potential. This current generates non-zero magnetic field which is then measured at anticipated moment in time determined by the time necessary for magnetic field to travel the distance from the SR to the detector.  b) Aharonov-Bohm experiment with single electron. The imposed superposition of trajectories ($e_L,e_R$) of electron 
    results in its entanglement with the magnetic field generated by the electron,  which finally leads to the entanglement with solenoid: $(\ket{e_L,B\odot, S_L}+\ket{e_R,B\otimes,S_R})/\sqrt{2}$. Here $B\odot,B\otimes$ represent states of the magnetic field generated by the moving electron and $S_{L,R}$ -- the states of the solenoid. This entanglement  allows for formulating Vaidman's loophole.}
    \label{Scheme}
\end{figure}
\subsection{Setup and experiment}
\begin{figure} 
     \centering
\includegraphics[width=0.8\columnwidth]{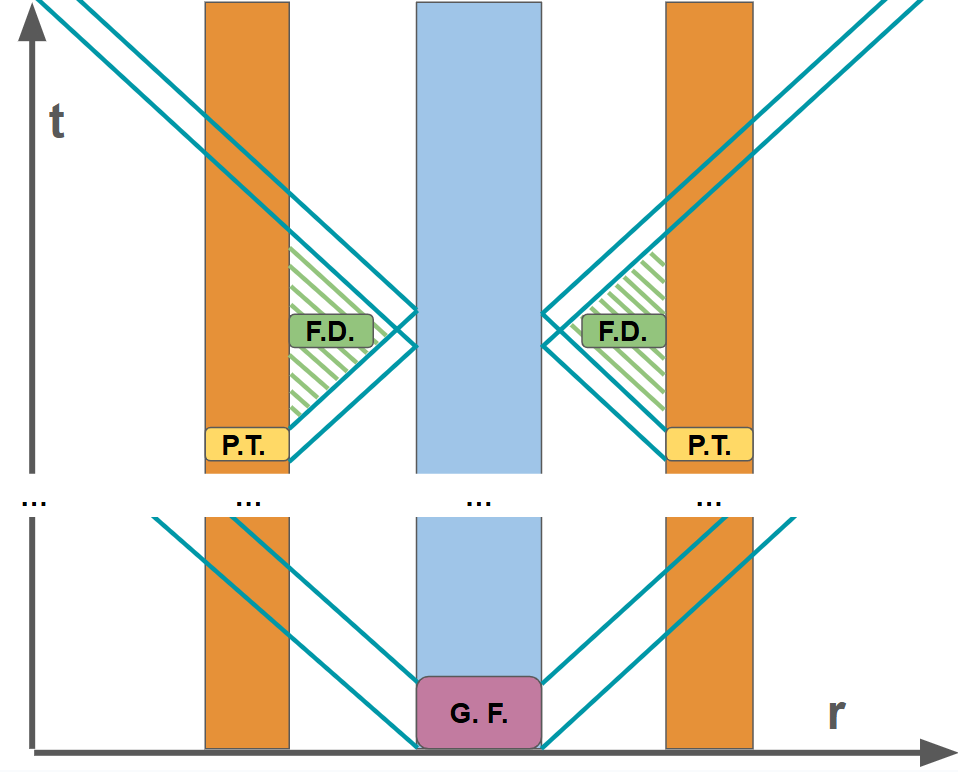}
        \caption{A space-time diagram of the crucial steps of our gedanken experiment. $r$ denotes the radial dimension of the system. Blue rectangle represents the cross-section of the solenoid, whereas orange rectangles represent cross-section of the ring. Blue solid lines represent light trajectories, which determine causal structure of the experiment. The experiment starts with the process of flux generation (purple rectangle \textbf{G. F.}), which includes turning on the current in the solenoid and is followed by dissipation of any fields induced in the ring by the initial impulse. The initial flux is \textbf{not equal} to the multiple of the flux quantum.
        The dissipation stage lasts sufficiently long in order to achieve a state of constant flux within the ring and lack of any electromagnetic fields propagating in between the ring and the solenoid, which is represented by the dots "..." . 
        After the dissipation stage a phase transition to the superconducting state (\textbf{P. T.}) is induced in the ring. Since a superconducting ring can surround only a quantized flux, and the information about non-quantized flux is accessible to the ring solely via vector potential $\vec A$,  a supercurrent is induced in the (now) superconducting ring, which forces flux quantization within the ring. 
        Finally just after the phase transition in the ring, the signal due to the field induced by the supercurrent reaches the magnetic field detector (placed anywhere within the hatched green triangles \textbf{F.D.}) where it is measured. Detector's  time window is chosen such that the only source of the signal after phase transition could be from the ring side. Then induction of the field in the system must have been caused by vector potential "informing" the superconducting ring that the flux surrounded by the ring is not properly quantized.   
        }
        \label{fig:spacetime}
\end{figure}

Let us start with introducing a setup for which one can make a prediction, using standard theory of superconductivity, of an event which appears to be unexplainable by solely locally interacting electromagnetic fields.
Consider  a setup that consists of a superconducting ring and centered inside the ring infinitely long solenoid (one could also imagine setup  with toroidal solenoid), see Figure \ref{Scheme} a) and \ref{fig:spacetime}. The solenoid and the ring are separated by some substantial distance $d$, the role of which we will discuss later on. Let us first present the particular steps of the experiment and then discuss their individual implications.

\begin{figure} 
     \centering
\includegraphics[width=0.99\columnwidth]{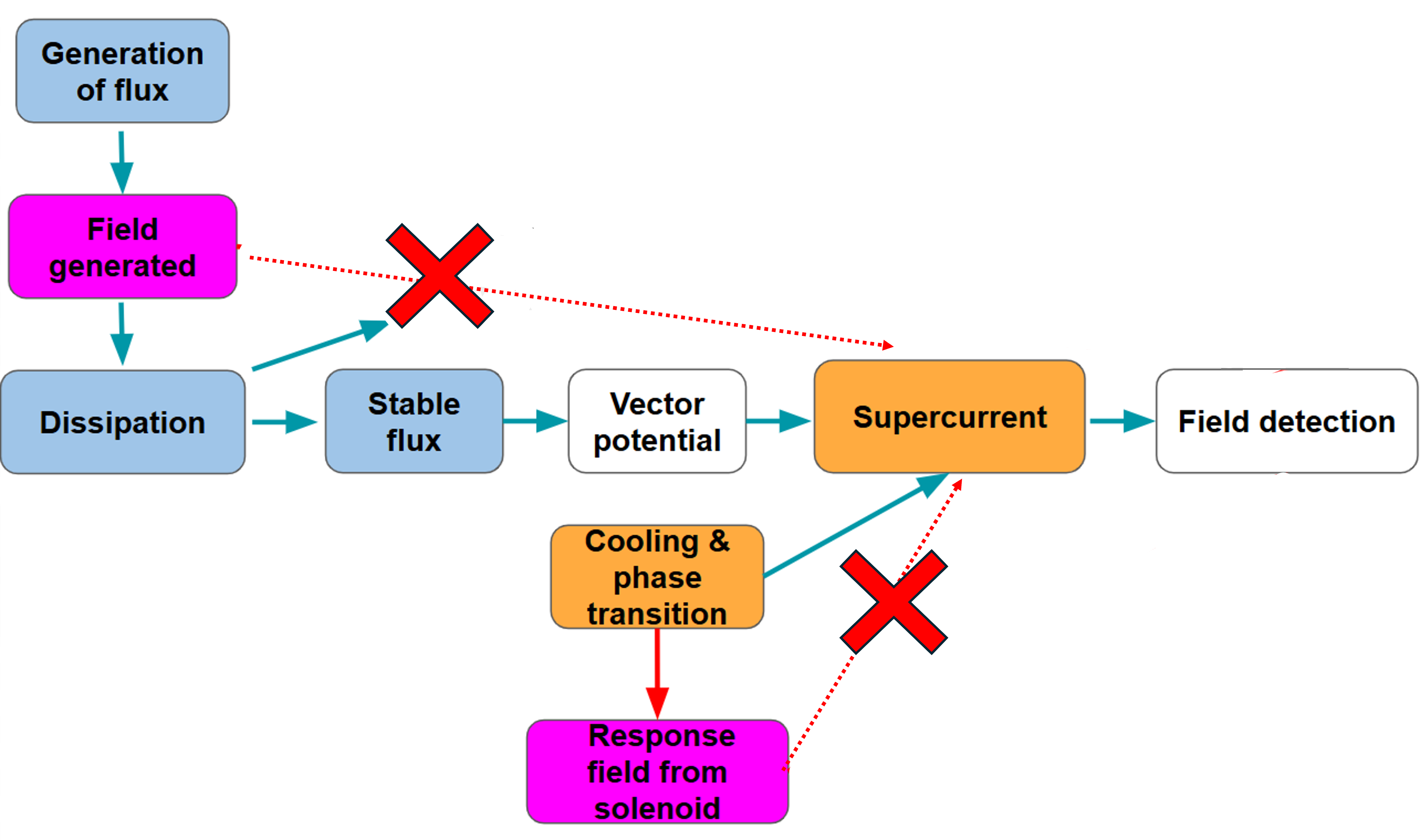}
        \caption{Causal diagram representing proposed experiment. Blue rectangles represent events taking place within the solenoid, orange ones -- within the ring, and white ones - in between the ring and the solenoid. Green arrows represent the desired causal structure explaining observation of a field between ring and the solenoid, in which the existence of a vector potential is a necessary factor for inducing the super current compensating lack of flux quantization within the superconducting ring after the phase transition. Purple rectangles and red arrows represent possible loopholes in the experiment, namely causal explanations for observing magnetic field in between the ring and the solenoid, which do not demand the vector potential as a part of a causal explanation of the final effect. The first loophole due to initial generation of the flux is suppressed by dissipation process, leading to the stable flux (this is denoted by a dotted red arrow). Second loophole is due to possible  field generation during the process of cooling which leads to phase transition. This could then provoke generation of response field from the solenoid to which one could try to attribute generation of the supercurrent. This loophole is rejected by appropriately designed temporal structure of the experiment, namely by performing field measurement precisely in the green-hatched spacetime region as shown in Fig. \ref{fig:spacetime}.}
        \label{fig:causal}
\end{figure}

One considers the initial state of the system in which the ring is still in a normal state and there is no flux of magnetic field through the solenoid. Then one proceeds as follows: 
\begin{enumerate}
    \item One switches on the current in the solenoid to achieve some initial target magnetic flux $\Phi_0$ which is \textit{not} equal to the multiple of flux quanta $n\pi\hbar/e$ where $n\in \mathbb{N}$.
    \item One awaits stabilization of the whole system and thus also of  the flux in the solenoid. 
    \item One couples the ring to the cold bath and symmetrically cools the ring to the critical temperature of the superconductor $T_c$ in time shorter than light needs to traverse the distance $2d$.
    \item The response of the field is registered by a sensor placed in between the ring and the solenoid,  in the region of spacetime where no response from solenoid to cooling procedure could be present.
\end{enumerate}
After phase transition occurring in the third step, according to the London equation, a supercurrent appears in the ring in response to the presence of a nontrivial vector potential related with the flux confined in the solenoid. In the absence of the field, this supercurrent has the task of quantizing the flux inside the ring. 
Thus, as a result, magnetic field should appear in the system that would change the flux through the ring towards the multiple of the flux quanta anticipated for the given initial $\Phi_0$. One then can measure the field in the spacetime region where only the response to cooling procedure from the ring could be present, ensuring the source of this field (ring).  Therefore, if at the moment at which the signal from the superconductor after phase transition should arrive to the field detector one registers anomalous magnetic field,  this certifies that either potentials are more fundamental than fields or that fields themselves admit nonlocal interactions. It is important to note that it is the response at the specific point in time that gives this conclusion and not necessarily observing the flux quantization. Figure \ref{fig:causal} presents the causal diagram of the experiment including all possible justifications of the final effect. In the next section, we present this argument in more detail and describe the role of each step of the experiment. 

\subsection{Discussion of the steps of the experiment}
It is an important matter when theoretically discussing fundamental aspects of physics to ensure that, in principle, one is physically able to prepare the initial state for the experiment in a continuous manner  (i.e., system evolves without unphysical discontinuous jumps of values of physical quantities) and to take into account the state preparation procedure during the reasoning. In other words, one wants to avoid physically unmotivated discontinuous quenches in the discussion.  This is because otherwise one can always argue that if a given system is prepared in some exotic initial state using such a quench, then the observed effects could be imprinted to the state on the level of the state preparation during the real physical processes leading to this effective quench. The standard discussion of the flux quantization and the AB effect often refers to the static scenario, in which  the flux in the solenoid is already established. However, the initial stage of turning on the current in the solenoid is a dynamic and gradual process (even if sudden) in which a non-zero field appears outside the solenoid. This can be seen directly from the Faraday's law: $\oint d\vec l\cdot \vec E=-\frac{d\Phi}{dt}$. Therefore, we start our considerations from a simple physical state where we have no flux $\Phi$ through the solenoid, and the first three steps simply build the ground for the effect of flux quantization to occur. This is done in such a way that any impact of a non-zero field appearing outside the solenoid is removed with minimal assumptions on the underlying mechanisms. As a side remark note that ignoring the continuous character of varying physical quantities in the context of electrodynamics can lead to paradoxical conclusions, as in the famous case of an apparent non-causal pre-acceleration of point charged particle being accelerated by an  external force \cite{Rohrlich97}. Namely the classical equation of motion for a charged particle, the so-called Abraham-Lorentz equation, contains the \textit{radiation reaction} term, which is proportional to the time derivative of acceleration of the particle. Now, if the external force applied to the particle varies in time too quickly (the extremal example being a step function-type behavior), the particle apparently starts to accelerate \textit{before} the force is applied. This apparent effect violating causality is a consequence of a nonphysical discontinuity in the time dependence of the external force; see the excellent analysis done by Yaghjian \cite{Yag07}.

The first step simply introduces the field into the system. We require that the flux is not a multiple of flux quanta to be able to observe flux quantization in the end. Then we await the stabilization in the second step as the initial impulse can provoke some response from the ring. Still, the response of the electrons in the ring that could impact the flux through the solenoid is suppressed with time, as the ring is in a normal state and any currents appearing in it will be damped. In the end, we are left with the vector potential $\vec A_0$ corresponding to the absence of the field in the region of the ring with the field confined only to the solenoid. In the third step  by the phase transition we introduce a necessary element for the flux quantization,  that is a superconducting ring. When reaching the critical temperature, we arrive at the configuration that will dynamically lead to flux quantization as a response of superconducting ring to the vector potential $\vec A_0$.

Let us here analyze some key points  concerning these steps. In the second step, one can see the system also as coupled to some thermal bath and the stabilization of the system is simply corresponding to thermalization of the  electronic state of the ring. In this process, any information about the initial impulse is dissipated in the bath and irreversibly lost. Here, it is crucial that the ring is in the normal state. Otherwise, one would get an immediate response from the ring in the form of the supercurrent, and one would not be able to distinguish whether the resulting flux quantization comes from the interaction with the potential or from some unspecified interaction with the field.  The normal state of the ring ensures that there are no persisting currents that could locally store the information about the flux in the solenoid, and there is obviously no response of the ring to the vector  potential $\vec A_0$. More specifically the state of the electrons in a normal metal is quickly thermalized  by the electron-phonon interaction (of order of picoseconds), which comes from scattering of electrons with the lattice \cite{Allen}. When electron-phonon coupling is weak (low-temperature regime) this could in principle be described by Markovian Lindblad master equation \cite{Petruccione,Lidar} for electron density matrix $\hat\rho$:
\begin{equation}
\frac{d\hat\rho}{dt} = -\frac{i}{\hbar} [\hat H_S+\hat H_{LS}, \hat\rho] + \sum_{\omega} \sum_{i} \gamma_{i}(\omega) \left( \hat L_{i}(\omega) \hat \rho \hat L_{i}^\dagger(\omega) - \frac{1}{2} \left\{\hat L_{i}^\dagger(\omega) \hat L_{i}(\omega),\hat \rho \right\} \right), 
\end{equation}
where $\hat H_S$ is the Hamiltonian describing electronic energy levels in bands, $\hat H_{LS}$ is the Lamb-shift due to interaction with the bath, and $\hat L_{i}(\omega)$ are system jump operators that correspond to the decoherence in the system of electrons in the form of incoherent transitions with Bohr frequency $\omega$ induced by the interaction with the bath with transition rates $\gamma_i(\omega)$. For example, assuming 1D almost free electron model of metal \cite{Kittel} one can  find jump operators from the coupling Hamiltonian between system and the bath,  which for electron-phonon interaction in general reads \cite{EL-phonon} $\sum_{k,q,\sigma}g(k,q)\hat c^\dagger_{k+q,\sigma} \hat c_{k,\sigma}(\hat b_q+\hat b_{-q}^\dagger)$, where $g(k,q)$ is the coupling constant, $\hat c_{k,\sigma}$ is the annihilation operator of an electron in Bloch energy eigenstate with wave-vector $k$  and spin $\sigma$ and $\hat b_q$ denotes the annihilation operator of a phonon with wave-vector $q$.  Note that the description by the Lindblad equation does not allow for proper full description of ultra-fast non-equilibrium processes \cite{Beyond-TTM,Beyond-TTM2,Beyond-TTM3} due to their non-Markovian character and because of the small energy gaps between electronic eigenstates in the band for which secular approximation does not hold. However, on sufficiently long timescales secular Markovian description becomes valid \cite{Time-scales}.  For phononic bath in thermal equilibrium the KMS condition $\frac{\gamma_{i}(\omega)}{\gamma_{i}(-\omega)} = e^{\hbar \omega / k_B T}$ is fulfilled which leads to detailed balance \cite{Lidar}. Then under the assumption of ergodicity of the system the thermal Gibbs state becomes an attractor and the  density matrix asymptotically converges to it:
\begin{equation}
\hat\rho_{th}=\frac{e^{-\beta \hat H_S}}{\text{tr}[e^{-\beta \hat H_S}]} ,
\end{equation}
where $\beta$ is the inverse temperature. As this state is fully determined by the Hamiltonian, any information about an initial impulse is removed from the state of electrons by this process.\footnote{Note that the Gibbs state could not be exactly a steady state if we consider interactions with different baths that have different temperatures, like for example coupling of electrons to photons when the ring is considered to be in a vacuum, which represents a bath of a different temperature from the phononic one. While such situations are still active research field, generally
this can still lead to a non-equilibrium steady state (due to heat flow), that is  independent of the initial state \cite{NESS1,Mulit-bath}. However,  because the dominant interaction in considered system is the electron-photon interaction, this deviates the steady state marginally.}   Still, in theory the system could be not fully ergodic due to some symmetries that can lead to decoherence-free subspaces \cite{DFS1,DFS2,DFS3}, and this could prevent full thermalization. Still these symmetries can be broken by defects and different interactions, removing decoherence-free subspaces, with the effective result being that thermalization of such subspace can be slow down. Additionally, after the electron-phonon interactions while the information about impulse for electronic system seems to be irreversibly lost, it is still contained in the region of the ring as phonons are confined to it and still some small memory effects can in principle be present making small deviations from thermal state. This makes the third step of the cooling even more essential to the argument.

The third step additionally ensures spatial removal of the information about the initial impulse from the ring. The cooling can be performed in multiple steps. Then this is equivalent to taking two uncorrelated cold heat baths. First, one cools the ring with one bath that removes any remaining information about the initial impulse from the ring as it is transferred into the bath with dissipated energy (analogously as in discussed electron-phonon interaction). This ends the dissipation of the initial impulse, and the only remaining local source of information about the flux $\Phi_0$ confined to the solenoid is the vector potential $\vec A_0$. After this initial cooling, one decouples the ring from the first cold bath and couples it to the second one that is used for the phase transition. One could argue that some information about the impulse is left in the ring after this process, as it disappears completely only asymptotically, and the thermalization is never full. However, note that cooling can be performed using different methods or their combinations. As different methods of cooling remove information from the system in different ways, leading to different decoherence-free subspaces (if any), and thus also different steady near thermal states, this epsilon information left should also have a different character, and thus if it would cause the response of the superconducting ring one would expect some significant differences in the signal.  More importantly, for each different way of storing information about impulse, there would have to be a separate mechanism for transforming this information into a signal that has the same effect in the end.  Therefore, persisting on this point, in fact, is equivalent to postulating existence of a large class of unknown mechanisms with the additional assumption that all of them are solely explainable with local field interactions.  When one obtains the expected result that the signal is the same for different cooling methods, then it is even harder to justify such models, as then they have to additionally converge to the same response on the same time scale. 

In the third step, we consider that the cooling is done in the time shorter than light needs to travel from the ring to the solenoid and back to the ring. This step is to ensure that if in some way the cooling process generates some field, the possible response from the solenoid cannot affect in any way the field detection performed in between the ring and the solenoid (hatched green region in the Figure \ref{fig:spacetime}). Such a space-time configuration of the field measurement removes the possibility that the detected field change has been caused by the interaction of the response impulse from the solenoid with the superconducting ring.

Finally after the phase transition one arrives at a situation in which the flux is not quantized in the solenoid but we have a superconducting ring around it. We know from experiment and theory of superconductivity that flux through the superconducting ring has to be quantized. Of course, here these two statements do not provide any paradox but simply imply that we have to dynamically evolve our system to a state that involves quantized flux. Of course, at the moment of phase transition the solenoid does not have any information about this event, so it does not respond immediately to it. However, the region of the ring has information about the flux in the form of a vector potential $\vec A_0$. This potential, based on our arguments, is the only source of the information about the enclosed flux  (or eventually some other unknown physical entity, which in some approximation behaves as such potential). Then the ring has also the mechanism of extracting this information by the interaction of the Cooper pairs with vector potential, which results in the appearance of appropriate supercurrent. The effective dynamics of this process is described by the time-dependent Ginzburg-Landau equation that determines the time evolution of the  order parameter $\psi(\vec x,t)$ which is interpreted as a macroscopic wave function of superconducting electrons. Here, $|\psi(\vec x,t)|^2$ is interpreted as proportional to the local density of Cooper pairs, and $\varphi=\text{arg}(\psi(\vec x,t))$ is the local superconducting phase. One can find, using some approximations (see \ref{app_TDGL}), that one has a supported solution  for which for short times after the phase transition the amplitude is exponentially amplified, while for other solutions it is exponentially suppressed. This supported solution leads to the supercurrent:
\begin{equation}
    j(t) = \frac{2e\hbar}{m^* R}|\psi(t)|^2 \left( n - \frac{\Phi_0}{\Phi_Q} \right),\label{eq:supercurrent}
\end{equation}
where $\Phi_Q$ is the flux quantum, $m^*$ is the mass of charge carrier, $R$ is the radius of the superconducting ring,  and $n$  is an integer minimizing $\left( n - \frac{\Phi_0}{\Phi_Q} \right)^2$. This current generates magnetic field which has to compensate for the flux inside the ring such that the total flux within the ring becomes quantized. 
Whether the final flux increases or decreases depends on the initial flux $\Phi_0$ as the superconductor can increase the flux to the next multiple of flux quanta or decrease it to the previous one \cite{Kenawy18,time_to_Q}. 
 Detection of the signal from initially generated current would certify physicality of the potentials. This is because the signal emerges from a zero-field region with the information from the past field removed, and one needs another quantity to cause this signal, and in our case the potential $\vec A_0$ provides a sensible model for such a quantity. Of course, another option is that the interaction is non-local. 

Note that one does not suspect any signal from the ring whenever there is no flux through the solenoid. Thus, one can easily verify that the first signal in our experiment is not simply provoked solely by the phase transition itself, but it must be accompanied by the potential $\vec A_0$. Even if one observes some signal at phase transition for zero flux case, one can simply look for the anomaly from this signal in our scenario. 

Let us stress that in this reasoning, it is not the flux quantization which certifies the result but rather the appearance of the signal at correct time in the correct place. The flux quantization is, of course, the final result of the interaction and tells us that something must have happened in the time between phase transition and the final stabilized flux-quantized state. However, we need to certify the origin of the interaction and not its result. This is also why it is important to analyze the setup in the picture where there is no instantaneous interaction between the ring and solenoid. Taking all of that into account, one should perform measurement on the magnetic field to detect the signal in the region of spacetime where any response of the solenoid to the cooling process could not arrive (see Figure \ref{fig:spacetime}). 

Let us discuss this problem from another perspective.  As the London equation is an equilibrium equation, it is clear that its solution can be reached only after  the information about phase transition from one side of the ring reaches the other side of the ring. Thus, necessarily after this signal also reaches the solenoid, making it again unclear what is the source of this current. This explains why the flux quantization itself is not enough to state the  physicality of potentials. What is crucial for our argument is that opposite sides of the ring \textit{do not need} information about the phase transition on another side to generate the initial current.  To notice this, let us recall that wavefunctions of electrons in metal are generally delocalized across the bulk (Bloch theorem). Now, one can always decompose the delocalized wavefunction into a superposition of localized components, which are affected only by the local environment by the assumption of local interactions. However, these local changes in a global way affect the amplitude of this delocalized wavefunction. So the local environment will affect local parts of the wavefunction, e.g. by damping local parts of wavefunction unsupported by the effective Hamiltonian, thus also damping globally unsupported delocalized states that have high overlap with these localized wavefunctions and reversely for the supported states (see  \ref{app_local_nonlocal} for intuitive physical picture). 

\begin{figure}
\centering
\subcaptionbox{}
{\includegraphics[width=0.49\linewidth]{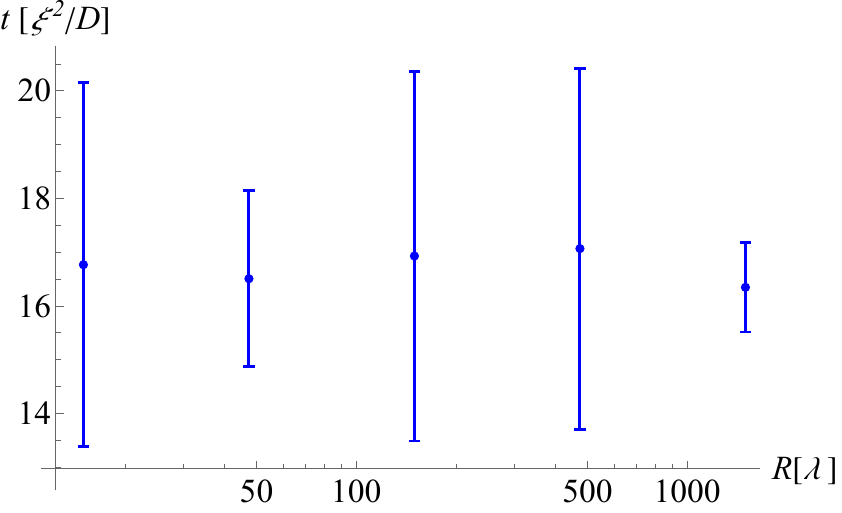}}
\subcaptionbox{}
{\includegraphics[width=0.49\linewidth]{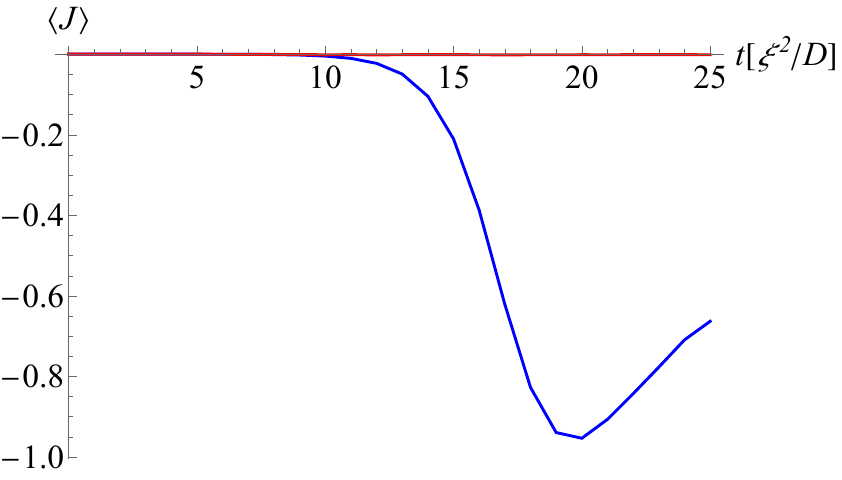}}
\caption{(a) Average time interval $t$ needed to achieve the asymptotic regime for different radii $R$ of the ring while keeping constant ratio between the initial flux $\Phi_0$ and radius $R$. Radius is expressed in the units of penetration depth $\lambda$ whereas time is presented in units  $\xi^2/D$, in which $\xi$ stands for coherence length and $D$ for diffusion coefficient. Results are obtained from 50 runs of simulation. Bars in the plot represent sample standard deviation. Clearly the time interval of achieving equilibration appears to be independent of the radius. Time interval needed for light to traverse one penetration depth is given by $D\lambda/c\xi^2$, with $c$ being the speed of light. This time can vary depending on the material and it's purity. For  Niobium  $\xi\approx\lambda\approx 4\cdot 10^{-8}$m and $D\sim10^{-4}$ for impure samples \cite{Diff} and for pure samples $D$ can reach $D\sim10^{-1}$ (based on estimation from mean free path $810$nm \cite{diff2}),  and thus this time is of the order $10^{-5}-10^{-2}\xi^2/D$ depending on the purity. Thus, in this example, to obtain the effect before signal reaches solenoid  one would need to choose a ring of radius of the order of  $10^7-10^4\lambda$, which is even in the worst case scenario in the reasonable range of decimeters. (b) Time dependence of average dimensionless current density $\langle J\rangle$ integrated over the ring  (see \ref{app:simulation} for details). Average was calculated over 50 runs of simulations. Blue curve was obtained for flux $\Phi_0=1000.2\, \Phi_Q$, whereas red curve for $\Phi_0=0$, and $R=1500\lambda$. Clearly nonzero flux results in organized current in the ring which however does not appear when there is no flux through the solenoid. While for this example $R$ is too small to guarantee that currents are big enough to be measurable before signal passes to solenoid, one should expect that this behavior extends also to wider rings. In  \ref{app:simulation} we present such behavior using simulations which utilize additional approximations. }\label{fig:time}
\end{figure}

Let us also notice that Ginzburg-Landau equations indeed allow for the formation of an initial organized current before the first information about phase transition reaches the solenoid. To this regard, we simulate the time interval needed to achieve $99\%$ of the amplitude of the order parameter corresponding to the asymptotic solution, for different radii $R$ of the ring and enclosed flux $\Phi_0$, while  keeping approximately constant ratio $\Phi_0/R$. The results are shown in Figure \ref{fig:time} a). We  observe that the time interval for reaching asymptotic regime is approximately constant. Therefore, one can find high enough $R$ such that the time  for appearance of initial structured currents is shorter than the time interval necessary for light to travel the distance between the ring and a solenoid. For more details on the presented simulation see \ref{app:simulation}. In Figure \ref{fig:time} b) we show the contrast of the behavior of the currents for the case with and without the initial flux within the solenoid.

Note that, by the construction of the experiment these claims are in principle falsifiable as in the opposite case there will be no signal at all in the space-time region chosen for performing the field measurement.   

\subsection{Vaidman's loophole}
While our argument shares some similarities with the argument based on the AB effect, there is, in fact, a crucial difference. Vaidman showed in the work \cite{Vaidman12} that the phase difference in the state of electrons in the AB effect can be seen as a result of the interaction of the solenoid with the magnetic field generated by the ``flying'' electron. This field is generated because in the AB experiment one imposes that an electron moves on one of  two ``trajectories'' that form a loop around the solenoid to finally perform an interference experiment (in \cite{Vaidman12} two half-circles were considered). This movement of the charge on the non-straight path results in the appearance of a magnetic field,  see Figure \ref{Scheme} b). In our scenario, there is no \textit{externally imposed} motion of the charge carriers (Cooper pairs).
In other words, if one removes the solenoid, there is no organized movement of charge carriers in our case (there is no flux at all), whereas in the AB effect, the electron has its predefined superposed paths regardless of the presence of the solenoid. This results in fundamentally different states of the system in the case of removed solenoid. For the AB  setup one has a maximally entangled state between the magnetic field generated by the moving electron and the electron's path degree of freedom: $(\ket{e_L,B\odot}+\ket{e_R,B\otimes})/\sqrt{2}$, in which the first term corresponds to the electron in the left arm and the second to the electron in the right arm and the field  has an opposite direction in these two cases. This entanglement is crucial for Vaidman's argument. However, in the case of the superconducting ring configuration, one has a separable state $\ket{\vec j=0,B=0}$ with zero current density $\vec j$ and zero magnetic field or at most small random local fluctuations. In other words, for the AB experiment when the solenoid is removed, one has a non-trivial magnetic field, while in our setup there is no magnetic field. This removes the argument that, in fact, the effect is due to the magnetic field generated by the charge carriers in the first place and not due to the interaction with the vector potential. This is because in our case the supercurrent is not imposed externally but is a result of the interaction with $\vec A_0$ and it simply does not appear otherwise.

\section{Conclusion}\label{sec13}
In this paper, we consider the gedanken experiment which gives a premise that either potentials are physical or field interactions are nonlocal. This experiment is based on the interaction of a superconducting ring with a vector potential which results in the flux quantization.
This approach differs from typically considered in this context Aharonov-Bohm effect, the reasoning of which was shown to have a loophole (Vaidman's loophole) in the context of proving physicality of potentials. Crucially we consider the superconducting ring to be significantly separated from the solenoid. This separation allows us for a non-standard analysis of the flux quantization, which is often described in the approximation of instantaneous interactions. This noninstantenous approach allows us to clearly distinguish cause from effect and therefore better understand the causal structure of the experiment. Another crucial point in the scheme is that one starts from the ring in the normal state, which  allows for the dissipation of any information about the field that originates from switching on the current in the solenoid. Therefore, after phase transition, the field generated by the ring has to originate from the interaction with the vector potential under the assumption of local interactions. 

Importantly, the proposed gedanken experiment is free from the Vaidman's loophole that nullified conclusions of original reasoning of Aharonov and Bohm. This is because there exists a fundamental difference between the AB experiment and the one based on flux quantization. In the first case, one deals with the entangled state of charge carriers with magnetic field generated by their movement, while in the former one there is a separable state. Note that sometimes more general class of problems, i.e., geometrical phase in electrodynamics is refereed to as AB effect. Then in such a view, it is argued that the flux quantization is a consequence of the AB effect. It might seem contradictory that we argue that the flux quantization helps to solve  the problem of potentials when the AB effect itself cannot. However, here we specifically refer to the AB effect as the effect of gaining phase difference between paths of the electron around a long solenoid and not in this more general meaning.

Still, our considerations do not give a definite answer, but only spotlight that the problem of physicality of  potentials still needs careful investigation. This is because to achieve our conclusion we used a well-established but still phenomenological model of time-dependent Ginzburg-Landau equations. It might be always the case that, in such a scenario the model for some reasons breaks in the regime of parameters and times needed to reach our conclusion. Therefore, further research which would refine the model is necessary. Especially interesting is to obtain quantitative predictions for the expected fields as the approximations used by us allow only for describing qualitative behavior of the system. This could then be a foundation for the feasibility assessment of potential experiment that could give us a definite answer to the problem. Finally, let us note that the concept used in this paper potentially could be also applied in the Aharonov-Bohm like-setup. This is because as noticed in \cite{Vedral20} the phase is acquired locally on the path of the electron and this could be tested by local tomography. Thus, one could try to measure this partial phase difference on the path of the electron around the solenoid before information about splitting to right and left paths reaches the solenoid.

\section*{Acknowledgements}

This work is supported by the IRAP/MAB programme, project no. FENG.02.01-IP.05-0006/23, financed by the MAB FENG program 2021-2027, Priority FENG.02, Measure FENG.02.01., with the support of the FNP (Foundation for Polish Science).

\appendix
\section{Initial current in experiment from time-dependent Ginzburg--Landau theory}\label{app_TDGL}
The dynamic properties of the superconductor can be described by  means of the time-dependent Ginzburg--Landau theory (TDGL). It describes the evolution of the superconducting order parameter \( \psi(\vec{r}, t) \) by means of the equation \cite{TDGL}: 
\begin{equation}
\Gamma^{-1} \left( \frac{\partial}{\partial t} + i\frac{2e}{\hbar} \phi \right) \psi
= -\alpha \psi - \beta |\psi|^2 \psi - \frac{1}{2m^*} \left( -i\hbar \nabla - 2e \vec{A} \right)^2 \psi,\label{app:GL}
\end{equation}
in which $\alpha = \alpha_0(T - T_c),$ where $T$ is the temperature, $T_c$ is the critical temperature, $\alpha_0,\beta > 0,$
whereas $\{\vec{A}, \phi\}$ are the electromagnetic potentials (in the notation we omit $\vec r$ and $t$ dependence for readability). Further, $e$ stands for electron charge, $m^*$ is an effective mass of charge carriers and $\Gamma=2mD/\hbar^2$ is a coefficient dependent on purity of the superconductor and $D$ is diffusion coefficient. The evolution of the order parameter allows to  calculate the evolution of the  supercurrent which is given in terms of $\psi$ as:
\begin{equation}
\vec{j}_s = \frac{2e}{m^*} \Re \left\{ \psi^* \left( -i\hbar \nabla - 2e \vec{A} \right) \psi \right\}.
\end{equation}
As a representation of the dynamical changes of the current under vector potential $\vec A$ in the considered setup, let us consider an approximate solution. As an approximation we assume the ring to be thin enough so that the vector potential is approximately constant across it. This reduces the dimensionality of the problem to only single polar coordinate $\varphi$. We also neglect the back-reaction of the field induced by the supercurrent, decoupling our equation from Maxwell's equations, as we are mostly interested in the existence of the initial impulse.

Consider an ansatz solution of the form of Fourier series component:
\begin{equation}
    \psi(\varphi, t) =  c_n(t) e^{i n \varphi},
\end{equation}
with $n\in\mathbb{Z}$ and $c_n(t)\in\mathbb{R}$. We assume a constant amplitude  over the ring, as our setup is assumed to be fully rotationally symmetric.  Choosing the vector potential in the system as: $\vec{A} = A_\varphi = \frac{\Phi}{2\pi R}\hat{\varphi} $, where $\Phi$ is the flux and $R$ is the radius of the ring, one can find that the current is given by:
\begin{equation}
    j_s(t) = \frac{2e\hbar}{m^* R} c_{n}(t)^2 \left( n - \frac{\Phi}{\Phi_Q} \right),\label{app_current}
\end{equation}
where $\Phi_Q$ is the flux quantum. Thus, an important dynamic parameter is $\rho_n(t)=c_n(t)^2$, which represents the density of Cooper pairs. From equation (\ref{app:GL}) one can get that:
\begin{equation}
    \frac{\partial}{\partial t}\rho_n(t)=2c_n(t)\frac{\partial}{\partial t}c_n(t)=2\Gamma(-\lambda_n-\beta\rho_n(t))\rho_n(t),\label{app:eq__diff_rho}
\end{equation}
where 
\begin{equation}
    \lambda_n=\left( \alpha + \frac{\hbar^2}{2m^* R^2} \left( n - \frac{\Phi}{\Phi_Q} \right)^2 \right).
\end{equation}
Assuming that initially just after the phase transition the order parameter is small with $\rho_n(0)=\epsilon\ll1$, one gets the solution to this differential equation:
\begin{equation}
 \rho_n(t)=   \frac{\lambda_n\epsilon}{-\beta  \epsilon+ e^{2\Gamma \lambda_n t}(\beta \epsilon+\lambda_n)}.\label{app_rho_t}
\end{equation}
This solution behaves differently depending on the sign of $\lambda_n$.  Let us first consider the short-time regime, where in the differential equation (\ref{app:eq__diff_rho}) one can omit the nonlinear part, as it has a marginal damping contribution due to the initial condition. Then the solution is obtained simply by putting $\beta=0$:
\begin{equation}
     \rho_n(t)= \epsilon e^{-2\Gamma\lambda_n t}\,\,\,\text{for}\,\, \Gamma t\ll1. 
\end{equation}
Clearly, for $\lambda_n>0$ we have exponential suppression of the solution. However, for $\lambda_n<0$ the solution is supported, and one observes exponential growth and thus also growth of the current associated with this solution. 

In the long time limit ($t\rightarrow\infty$ ), if $\lambda_n>0$ clearly from (\ref{app_rho_t}) one gets $\lim_{t\rightarrow\infty}\rho(t)=0$ while for $\lambda_n<0$ one gets typical steady state solution of time independent Ginzburg-Landau equation $-\lambda_n/\beta$.

Note that, as anticipated, to observe the supercurrent one needs to have a ring below the critical temperature as only in this case $\lambda_n$ can be negative due to the first term $\alpha<0$. However, just after the phase transition where $|\alpha|$ is small, only $n=n_0$ that minimizes $\left(n - \frac{\Phi}{\Phi_Q} \right)^2$ can result in $\lambda_n<0$ and thus only this solution can be supported. Note that if the flux is a multiple of flux quanta: $\Phi=n\Phi_Q$ then from (\ref{app_current}) there is no current from the supported solution, but otherwise it is non-zero and its direction is determined by $\text{sgn}\left(n_0 - \frac{\Phi}{\Phi_Q} \right)$. A more general solution of the form $\psi(\varphi, t) = \sum_n c_n(t) e^{i n \varphi}$ will also converge to  $-\frac{\lambda_{n_0}}{\beta} e^{i n_0 \varphi}$, and since the nonlinear part of evolution equation is fully damping, it cannot help supporting any other nonzero solutions.

\section{Non-local impact of local operations}\label{app_local_nonlocal}
Let us present a simple picture that illustrates that local ``environment'' has a non-local effect on the delocalized wavefunctions. Consider a delocalized wave function $\ket{\Psi}=\frac{1}{\sqrt{2}}(\ket{R}+\ket{L})$ of a single electron, which is a superposition of the electron going right and going left. Here $\ket{R}$, $\ket{L}$ represent localized wavepackets of right and left going electron respectively, where the average momentum in orthogonal directions is $\langle \vec p_\perp\rangle=0$. Now, if in the location of the right wave packet one switches on electric field in the direction orthogonal to the average momentum, one changes the effective Hamiltonian in such a way that it no longer supports states with $\langle \vec p_\perp\rangle=0$ in the right region. In  effect, the  amplitude for the delocalized $\ket{\Psi}$ (which is a global entity) drops asymptotically only due to the local interaction  from 1 to $\frac{1}{2}$ as the state tends to $\ket{\Psi'}=\frac{1}{\sqrt{2}}(\ket{R'}+\ket{L})$ with $\ket{R'}$ denoting the time-dependent wave function of an accelerating  right-going electron that has negligible overlap with $\ket{R}$, i.e. $\langle R|R'\rangle\approx0$. Similarly, if one switches on the  field on both sides, the initial state becomes fully unsupported, and the amplitude of this state drops to $0$. However, the left side does not need to communicate to the right side that now states of the form $\ket{\Psi}=\frac{1}{\sqrt{2}}(\ket{R}+\ket{L})$ are unsupported, and instead the system favors states of the form  $\ket{\Psi'}=\frac{1}{\sqrt{2}}(\ket{R'}+\ket{L'})$ to start making changes to the amplitude of the initial state.   Translating this to our considerations, different parts of the ring do not need to communicate about the phase transition to start amplifying some delocalized states, because local interactions allow one to change the amplitudes of delocalized states, and as we are considering electrons in the ring modes that are delocalized from the beginning. To start with a system that has a desired supported solutions, the local actions just need to be done in synchronized way,  which is assumed in our reasoning. Otherwise, one would need to for example introduce position dependent $\alpha$ if the cooling is not uniform, and this could change the globally supported solution.

\section{Simulation of a time interval necessary to reach asymptotics}
\label{app:simulation}
To perform simulations we use normalized version of Ginzburg-Landau equations. We choose to normalize flux to flux quanta,  distance to penetration depth $\lambda$ and time to  $\xi^2/D$ where $\xi$ stands for coherence length and $D$ for diffusion coefficient of electrons in the material. We also choose to normalize the order parameter to $\tilde\psi(\varphi,t)=\sqrt{\frac{\beta}{|\alpha|}}\psi(\varphi,t)$. Ginzburg-Landau equation expressed with normalized quantities then reads \cite{Kenawy18}:
\begin{equation}
 \frac{\partial}{\partial t} \tilde\psi
= (1 -  |\tilde\psi|^2) \tilde\psi -  \left( -i\frac{1}{\tilde R\kappa}\frac{\partial}{\partial\varphi} - \tilde A \right)^2 \tilde\psi,\label{app:GL}
\end{equation}
where $\kappa=\lambda/\xi$ and $\tilde A=\frac{\tilde\Phi}{\kappa \tilde R}$. From this normalized order parameter the dimensionless current can be calculated as:
\begin{equation}
    \tilde j_s=\frac{-i}{2\kappa \tilde R}\left(\tilde\psi^*\frac{\partial}{\partial\varphi}\tilde\psi-\tilde\psi\frac{\partial}{\partial\varphi}\tilde\psi^*\right)-\tilde A|\tilde\psi|^2.
\end{equation}
We additionally include the white noise term  $\eta(\varphi,t)$ in the evolution of $\tilde\psi$ which introduces  random complex noise on each step of numerical simulation. The noise $\eta(\varphi,t)$  is chosen based on normal distribution with standard deviation $\sigma=0.000001$. This random values for noise are drown in 200 equally spaced points and then the noise is interpolated between those points. We then perform simulation starting from meta-stable point $\tilde\psi(\varphi,0)=0$. For simulations, we have chosen $\kappa=0.8$. The radius and flux are chosen to follow $\tilde R=1.5(\sqrt{10})^{i}$ and $\tilde \Phi=\lceil(\sqrt{10})^{i}\rceil+0.2$ in order to check the behavior of the system every half of an order of magnitude, and keep approximately constant ratio $\tilde \Phi/\tilde R\approx1/3$. In  the Figure \ref{fig:time} (b) in the main text we present the current density integrated over the ring $J(t)=\int_0^{2\pi}\operatorname{d}\!\varphi\,\tilde{j}_s(\varphi,t)$ averaged over simulation runs.

Note that the most meaningful results from such simulations are those when currents are still relatively small in relation to $\tilde A$, as we do not update the vector potential. Still, the time in which the non-exact asymptotics is reached  bounds from above  the time in which first small organized currents appear. This is because, while changes in vector potential due to currents could slowdown the process of stabilization of the system, this can have significant impact only after the initial current already started to have  non-negligible values. 

\begin{figure}
\centering
\subcaptionbox{}
{\includegraphics[width=0.49\linewidth]{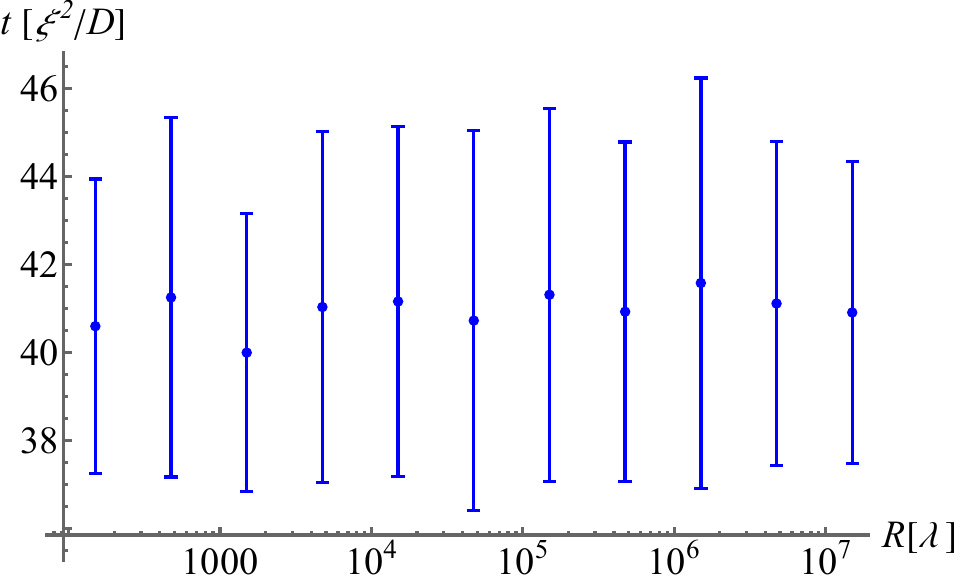}}
\subcaptionbox{}
{\includegraphics[width=0.49\linewidth]{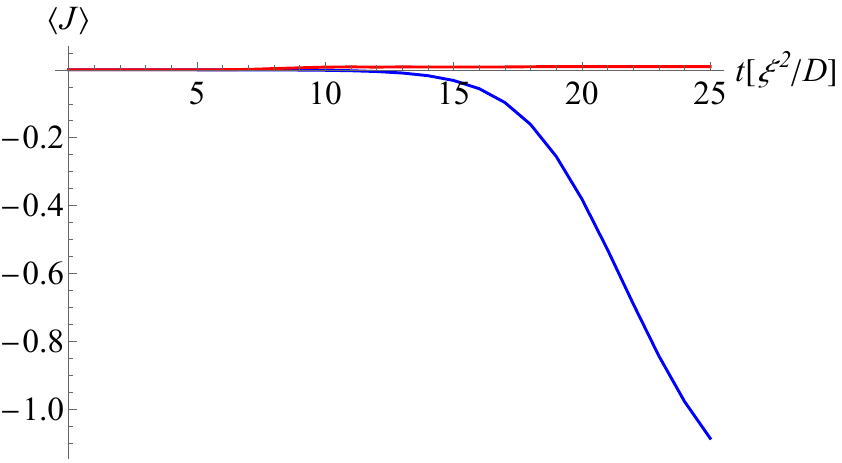}}
\caption{a) Time interval $t$ needed to achieve the asymptotic regime for different radii $R$ of the ring while keeping constant ratio between the initial flux $\Phi_0$ and radius $\tilde R$ for calculations with no enough dense grid to capture full solution. Bars in the plot represent sample standard deviation. For this case the time interval of achieving equilibration is also independent of the radius as in the main text. b) Time dependence of average dimensionless current $\tilde j_s$ integrated over the ring. Blue curve stands  for flux $\Phi_0=(10^7+0.2)\,\Phi_Q$, whereas red curve for $\Phi_0=0$ and $\tilde R=1.5*10^7$. Clearly nonzero flux results in organized current in the ring also for high value of $\tilde R$. The behavior for zero flux is also analogous to the simulation presented in the main text. Note that for this radius the time needed for light to pass to solenoid is of the order $10^2-10^5\xi^2/D$, and thus it is longer then our bound for the time needed for first non-negligible current to appear. }\label{fig:time_app}
\end{figure}

Note that, with increasing $\tilde\Phi$  solving the evolution equation to find a proper final state becomes infeasible as number of points in the grid needs to be considerably bigger than the magnitude of  $\tilde\Phi$ taken in the unit of flux quanta, in order to capture winding number of the solution. This limits the possible radius for which it is feasible to perform calculations. If number of points is too small the solution obtained numerically will be only an approximate slowly oscillating envelope of the real solution. Equilibration to this approximate solution has tendency to take more time, %overestimate time needed for equilibration
 what can be seen in the Figure \ref{fig:time_app}. Note that this still yields constant rate of equilibration, analogously to calculations with dense enough grid presented in the main text. This approach also allows us to make an approximate prediction about sufficiently big $R$ such that a measurable signal could appear before information on the phase transition arrives at the solenoid. 


\section*{Bibliography}

\begin{thebibliography}{10}

\bibitem{AB59}
Y.~Aharonov and D.~Bohm.
\newblock Significance of electromagnetic potentials in the quantum theory.
\newblock {\em Phys. Rev.}, 115:485--491, Aug 1959.

\bibitem{Aharonov15}
Yakir Aharonov, Eliahu Cohen, and Daniel Rohrlich.
\newblock {Comment on ``Role of potentials in the Aharonov-Bohm effect''}.
\newblock {\em Phys. Rev. A}, 92:026101, Aug 2015.

\bibitem{Allen}
Philip~B. Allen.
\newblock Theory of thermal relaxation of electrons in metals.
\newblock {\em Phys. Rev. Lett.}, 59:1460--1463, Sep 1987.

\bibitem{BCS}
J.~Bardeen, L.~N. Cooper, and J.~R. Schrieffer.
\newblock Theory of superconductivity.
\newblock {\em Phys. Rev.}, 108:1175--1204, Dec 1957.

\bibitem{berry1984quantal}
Michael~Victor Berry.
\newblock Quantal phase factors accompanying adiabatic changes.
\newblock {\em Proceedings of the Royal Society of London. A. Mathematical and Physical Sciences}, 392(1802):45--57, 1984.

\bibitem{BBBB}
Iwo Białynicki-Birula and Zofia Białynicka-Birula.
\newblock {\em Quantum Electrodynamics}.
\newblock Pergamon Press, Oxford, 1975.

\bibitem{Bocchieri78}
P.~Bocchieri and A.~Loinger.
\newblock {Nonexistence of the Aharonov-Bohm effect}.
\newblock {\em Il Nuovo Cimento A (1965-1970)}, 47(4):475--482, Oct 1978.

\bibitem{Bocchieri80}
P.~Bocchieri, A.~Loinger, and G.~Siragusa.
\newblock {The role of the electromagnetic potentials in quantum mechanics. The marton experiment}.
\newblock {\em Il Nuovo Cimento A (1971-1996)}, 56(1):55--72, May 1980.

\bibitem{Petruccione}
Heinz-Peter Breuer and Francesco Petruccione.
\newblock {\em The Theory of Open Quantum Systems}.
\newblock Oxford University Press, 01 2007.

\bibitem{Chambers60}
R.~G. Chambers.
\newblock Shift of an electron interference pattern by enclosed magnetic flux.
\newblock {\em Phys. Rev. Lett.}, 5:3--5, Jul 1960.

\bibitem{TD5}
S.~Rai Choudhury and Shobhit Mahajan.
\newblock {Direct calculation of time varying Aharonov-Bohm effect}.
\newblock {\em Physics Letters A}, 383(21):2467--2471, 2019.

\bibitem{Cohen2019}
Eliahu Cohen, Hugo Larocque, Fr{\u{A}}{\textcopyright}d{\u{A}}{\textcopyright}ric Bouchard, Farshad Nejadsattari, Yuval Gefen, and Ebrahim Karimi.
\newblock {Geometric phase from Aharonov-Bohm to Pancharatnam-Berry and beyond}.
\newblock {\em Nature Reviews Physics}, 1(7):437--449, Jul 2019.

\bibitem{combescot2022superconductivity}
R.~Combescot.
\newblock {\em Superconductivity: An Introduction}.
\newblock Cambridge University Press, Cambridge, 2022.
\newblock p. 13.

\bibitem{DEAVER1982178}
Bascom~S. Deaver and Gordon~B. Donaldson.
\newblock {An experimental demonstration of winding number dependence of the Aharonov-Bohm effect}.
\newblock {\em Physics Letters A}, 89(4):178--180, 1982.

\bibitem{Flux_q_exp1}
Bascom~S. Deaver and William~M. Fairbank.
\newblock Experimental evidence for quantized flux in superconducting cylinders.
\newblock {\em Phys. Rev. Lett.}, 7:43--46, Jul 1961.

\bibitem{Flux_q_exp2}
R.~Doll and M.~N\"abauer.
\newblock Experimental proof of magnetic flux quantization in a superconducting ring.
\newblock {\em Phys. Rev. Lett.}, 7:51--52, Jul 1961.

\bibitem{Ehrenberg49}
W~Ehrenberg and R~E Siday.
\newblock {The Refractive Index in Electron Optics and the Principles of Dynamics}.
\newblock {\em Proceedings of the Physical Society. Section B}, 62(1):8, jan 1949.

\bibitem{EL-phonon}
Feliciano Giustino.
\newblock Electron-phonon interactions from first principles.
\newblock {\em Rev. Mod. Phys.}, 89:015003, Feb 2017.

\bibitem{TDGL}
L~P Gor'kov and N~B Kopnin.
\newblock Vortex motion and resistivity of type-ll superconductors in a magnetic field.
\newblock {\em Soviet Physics Uspekhi}, 18(7):496, jul 1975.

\bibitem{Greenberger81}
Daniel~M. Greenberger.
\newblock {Reality and significance of the Aharonov-Bohm effect}.
\newblock {\em Phys. Rev. D}, 23:1460--1462, Mar 1981.

\bibitem{NESS1}
J.-T. Hsiang and B.L. Hu.
\newblock Nonequilibrium steady state in open quantum systems: Influence action, stochastic equation and power balance.
\newblock {\em Annals of Physics}, 362:139--169, 2015.

\bibitem{TD4}
Jian Jing, Yu-Fei Zhang, Kang Wang, Zheng-Wen Long, and Shi-Hai Dong.
\newblock {On the time-dependent Aharonov–Bohm effect}.
\newblock {\em Physics Letters B}, 774:87--90, 2017.

\bibitem{Kang22}
Kicheon Kang.
\newblock {Gauge invariance of the local phase in the Aharonov-Bohm interference: Quantum electrodynamic approach}.
\newblock {\em Europhysics Letters}, 140(4):46001, nov 2022.

\bibitem{Kenawy18}
Ahmed Kenawy, Wim Magnus, and Bart Sor{\'e}e.
\newblock {Flux Quantization and Aharonov-Bohm Effect in Superconducting Rings}.
\newblock {\em Journal of Superconductivity and Novel Magnetism}, 31(5):1351--1357, May 2018.

\bibitem{Kittel}
C.~Kittel.
\newblock {\em Introduction to Solid State Physics}.
\newblock Wiley, 2004.

\bibitem{Klein79}
U.~Klein.
\newblock {The inadmissibility of non-Stokesian vector potentials in quantum mechanics. Comments on a paper asserting the nonexistence of the Aharonov-Bohm effect}.
\newblock {\em Lettere al Nuovo Cimento (1971-1985)}, 25(2):33--37, May 1979.

\bibitem{time_to_Q}
Alvin~L. Kwiram and Bascom~S. Deaver.
\newblock Observations of the establishment of the quantized flux state in times as short as ${10}^{\ensuremath{-}5}$ sec.
\newblock {\em Phys. Rev. Lett.}, 13:189--190, Aug 1964.

\bibitem{TD1}
B.~Lee, E.~Yin, T.~K. Gustafson, and R.~Chiao.
\newblock {Analysis of Aharonov-Bohm effect due to time-dependent vector potentials}.
\newblock {\em Phys. Rev. A}, 45:4319--4325, Apr 1992.

\bibitem{Diff}
A.~Leo, G.~Grimaldi, R.~Citro, A.~Nigro, S.~Pace, and R.~P. Huebener.
\newblock Quasiparticle scattering time in niobium superconducting films.
\newblock {\em Phys. Rev. B}, 84:014536, Jul 2011.

\bibitem{DFS1}
D.~A. Lidar, I.~L. Chuang, and K.~B. Whaley.
\newblock Decoherence-free subspaces for quantum computation.
\newblock {\em Phys. Rev. Lett.}, 81:2594--2597, Sep 1998.

\bibitem{Lidar}
Daniel~A. Lidar.
\newblock Lecture notes on the theory of open quantum systems, 2020.

\bibitem{DFS2}
Daniel~A. Lidar and K.~Birgitta~Whaley.
\newblock {\em Decoherence-Free Subspaces and Subsystems}, pages 83--120.
\newblock Springer Berlin Heidelberg, Berlin, Heidelberg, 2003.

\bibitem{london1950superfluids}
F.~London.
\newblock {\em Superfluids}, volume~I.
\newblock Wiley, New York, 1950.
\newblock p. 152.

\bibitem{London1935TheEE}
Fritz London and Heinz London.
\newblock The electromagnetic equations of the supraconductor.
\newblock {\em Proceedings of The Royal Society A: Mathematical, Physical and Engineering Sciences}, 149:71--88, 1935.

\bibitem{TD3}
James Macdougall and Douglas Singleton.
\newblock {Stokes' theorem, gauge symmetry and the time-dependent Aharonov-Bohm effect}.
\newblock {\em Journal of Mathematical Physics}, 55(4):042101, 04 2014.

\bibitem{Vedral20}
Chiara Marletto and Vlatko Vedral.
\newblock {Aharonov-Bohm Phase is Locally Generated Like All Other Quantum Phases}.
\newblock {\em Phys. Rev. Lett.}, 125:040401, Jul 2020.

\bibitem{Rev_flux}
S.~Olariu and I.~Iovitzu Popescu.
\newblock The quantum effects of electromagnetic fluxes.
\newblock {\em Rev. Mod. Phys.}, 57:339--436, Apr 1985.

\bibitem{Beyond-TTM2}
Shota Ono.
\newblock Thermalization in simple metals: Role of electron-phonon and phonon-phonon scattering.
\newblock {\em Phys. Rev. B}, 97:054310, Feb 2018.

\bibitem{Osakabe86}
Nobuyuki Osakabe, Tsuyoshi Matsuda, Takeshi Kawasaki, Junji Endo, Akira Tonomura, Shinichiro Yano, and Hiroji Yamada.
\newblock {Experimental confirmation of Aharonov-Bohm effect using a toroidal magnetic field confined by a superconductor}.
\newblock {\em Phys. Rev. A}, 34:815--822, Aug 1986.

\bibitem{Pearle2}
Philip Pearle and Anthony Rizzi.
\newblock {Quantized vector potential and alternative views of the magnetic Aharonov-Bohm phase shift}.
\newblock {\em Phys. Rev. A}, 95:052124, May 2017.

\bibitem{Pearle1}
Philip Pearle and Anthony Rizzi.
\newblock {Quantum-mechanical inclusion of the source in the Aharonov-Bohm effects}.
\newblock {\em Phys. Rev. A}, 95:052123, May 2017.

\bibitem{Beyond-TTM3}
Ulrike Ritzmann, Peter~M. Oppeneer, and Pablo Maldonado.
\newblock Theory of out-of-equilibrium electron and phonon dynamics in metals after femtosecond laser excitation.
\newblock {\em Phys. Rev. B}, 102:214305, Dec 2020.

\bibitem{Rohrlich97}
F.~Rohrlich.
\newblock The dynamics of a charged sphere and the electron.
\newblock {\em American Journal of Physics}, 65(11):1051--1056, 11 1997.

\bibitem{Roy80}
S.~M. Roy.
\newblock {Condition for Nonexistence of Aharonov-Bohm Effect}.
\newblock {\em Phys. Rev. Lett.}, 44:111--114, Jan 1980.

\bibitem{Ryder}
Lewis.~H. Ryder.
\newblock {\em {Quantum Field Theory}}.
\newblock Cambridge University Press, Cambridge, 1985.

\bibitem{AB_shielding}
Pablo~L. Saldanha.
\newblock Aharonov-casher and shielded aharonov-bohm effects with a quantum electromagnetic field.
\newblock {\em Phys. Rev. A}, 104:032219, Sep 2021.

\bibitem{Saldanha2021}
Pablo~L. Saldanha.
\newblock {Local Description of the Aharonov--Bohm Effect with a Quantum Electromagnetic Field}.
\newblock {\em Foundations of Physics}, 51(1):6, Feb 2021.

\bibitem{DFS3}
Konrad Schlichtholz and Marcin Markiewicz.
\newblock Relativistically invariant encoding of quantum information revisited.
\newblock {\em New Journal of Physics}, 26(3):033018, mar 2024.

\bibitem{TD2}
Douglas Singleton and Elias~C. Vagenas.
\newblock {The covariant, time-dependent Aharonov–Bohm effect}.
\newblock {\em Physics Letters B}, 723(1):241--244, 2013.

\bibitem{Mulit-bath}
Shuichi Tasaki and Taku Matsui.
\newblock {\em Fluctuation theorem, nonequilibrium steady states and maclennan-zubarev ensembles of a class of large quantum systems}, pages 100--119.
\newblock World Scientific, 2023.

\bibitem{diff2}
Edward Thoeng, Md~Asaduzzaman, Philipp Kolb, Ryan M.~L. McFadden, Gerald~D. Morris, John~O. Ticknor, Sarah~R. Dunsiger, Victoria~L. Karner, Derek Fujimoto, Tobias Junginger, Robert~F. Kiefl, W.~Andrew MacFarlane, Ruohong Li, Suresh Saminathan, and Robert~E. Laxdal.
\newblock Depth-resolved characterization of meissner screening breakdown in surface treated niobium.
\newblock {\em Scientific Reports}, 14(1):21487, Sep 2024.

\bibitem{Tonomura86}
Akira Tonomura, Nobuyuki Osakabe, Tsuyoshi Matsuda, Takeshi Kawasaki, Junji Endo, Shinichiro Yano, and Hiroji Yamada.
\newblock {Evidence for Aharonov-Bohm effect with magnetic field completely shielded from electron wave}.
\newblock {\em Phys. Rev. Lett.}, 56:792--795, Feb 1986.

\bibitem{Vaidman12}
Lev Vaidman.
\newblock {Role of potentials in the Aharonov-Bohm effect}.
\newblock {\em Phys. Rev. A}, 86:040101, Oct 2012.

\bibitem{Vaidman15}
Lev Vaidman.
\newblock {Reply to ``Comment on `Role of potentials in the Aharonov-Bohm effect' ''}.
\newblock {\em Phys. Rev. A}, 92:026102, Aug 2015.

\bibitem{wakamatsu2024revisitingcontroversytimedependentaharonovbohm}
Masashi Wakamatsu.
\newblock Revisiting the controversy over the time-dependent aharonov-bohm effect, 2024.

\bibitem{Beyond-TTM}
Lutz Waldecker, Roman Bertoni, Ralph Ernstorfer, and Jan Vorberger.
\newblock Electron-phonon coupling and energy flow in a simple metal beyond the two-temperature approximation.
\newblock {\em Phys. Rev. X}, 6:021003, Apr 2016.

\bibitem{Webb85}
R.~A. Webb, S.~Washburn, C.~P. Umbach, and R.~B. Laibowitz.
\newblock {Observation of $\frac{h}{e}$ Aharonov-Bohm Oscillations in Normal-Metal Rings}.
\newblock {\em Phys. Rev. Lett.}, 54:2696--2699, Jun 1985.

\bibitem{Weinberg1}
S.~Weinberg.
\newblock {\em The Quantum Theory of Fields. Volume 1. Foundations.}
\newblock Cambridge University Press, Cambridge, 1995.

\bibitem{Time-scales}
Marek Winczewski, Antonio Mandarino, Gerardo Suarez, Robert Alicki, and Micha\l{} Horodecki.
\newblock Intermediate-times dilemma for open quantum system: Filtered approximation to the refined weak-coupling limit.
\newblock {\em Phys. Rev. E}, 110:024110, Aug 2024.

\bibitem{Wu75}
Tai~Tsun Wu and Chen~Ning Yang.
\newblock Concept of nonintegrable phase factors and global formulation of gauge fields.
\newblock {\em Phys. Rev. D}, 12:3845--3857, Dec 1975.

\bibitem{Yag07}
A.D. Yaghjian.
\newblock {\em {Relativistic Dynamics of a Charged Sphere: Updating the Lorentz-Abraham Model}}.
\newblock {Springer International Publishing}, New York, 2023.

\end{thebibliography}
\end{document}